\documentclass[sigconf,10pt,natbib=true,anonymous=false]{acmart}
\usepackage{gensymb}
\usepackage{array}
\usepackage{algorithmic}
\usepackage{graphicx}
\usepackage{textcomp}
\usepackage{xcolor}
\usepackage{xspace}
\usepackage{enumitem}
\usepackage{url}
\usepackage{breakurl}
\usepackage{caption}
\usepackage{subcaption}
\usepackage{multirow}
\usepackage[normalem]{ulem}
\usepackage[algo2e,linesnumbered,ruled,lined,boxed]{algorithm2e}
\graphicspath{{figures/}}
\SetKwInput{KwInput}{Input}                
\SetKwInput{KwOutput}{Output}              

\newcolumntype{L}[1]{>{\raggedright\let\newline\\\arraybackslash\hspace{0pt}}m{#1}}
\newcolumntype{C}[1]{>{\centering\let\newline\\\arraybackslash\hspace{0pt}}m{#1}}
\newcolumntype{R}[1]{>{\raggedleft\let\newline\\\arraybackslash\hspace{0pt}}m{#1}}

\newcommand{\ie}{\textit{i.e.,}\xspace}
\newcommand{\eg}{\textit{e.g.,}\xspace}

\usepackage{comment}

\begin{document}

\setcopyright{none}

\title{
App-Based Performance Characterization of Cellular and Wi-Fi Networks in Dense Stadium Deployments
}


\settopmatter{printacmref=false}

\author{Hardani Ismu Nabil}
\authornote{Both authors contributed equally to this research.}
\email{hardani.ismu.n@gmail.com}
\affiliation{%
  \institution{Sebelas Maret University}
  \country{Indonesia}
}

\author{Muhammad I. Rochman}
\authornotemark[1]
\email{mrochman@nd.edu}
\affiliation{%
  \institution{University of Notre Dame}
  \country{USA}
}

\author{S. M. Haider Ali Shuvo}
\affiliation{%
  \institution{University of Notre Dame}
  \country{USA}
}

\author{Joshua Roy Palathinkal}
\affiliation{%
  \institution{University of Notre Dame}
  \country{USA}
}


\author{Monisha Ghosh}
\affiliation{%
  \institution{University of Notre Dame}
  \country{USA}
}



\renewcommand{\shortauthors}{H.I. Nabil, M.I. Rochman, \textit{et al.}}

\begin{abstract}

The concentration of 77,622 spectators during football games at Notre Dame Stadium creates an exceptionally demanding environment for wireless infrastructure. To handle this extreme user density, the stadium deploys concurrent multi-tier networks serving outdoor users: an enterprise 5/6~GHz Wi-Fi network with $\sim$900 outdoor Access Points (APs) alongside high-density multi-carrier 4G/5G networks powered by a neutral-host small-cell Distributed Antenna System (DAS) with up to 129 unique cell identifiers (PCIs) per operator. This study evaluates user-perceived performance and QoE across these networks using commercial smartphones to execute web browsing, WhatsApp messaging, and Instagram media posting workloads.
Our empirical results reveal that while cellular networks deliver strong peak downlink performance in an empty stadium, game-day crowd loads heavily strain uplink and latency performance, triggering a severe cellular ``uplink gap.'' Under Non-Standalone (EN-DC) anchor congestion, web browsing handshakes suffer a catastrophic 5,983~ms P90 Time-to-First-Byte (TTFB), and image upload failure rates climb to 46\%. Furthermore, while narrow low-band FDD channels (\eg n5) maintain robust channel quality during uploads, they exhibit a 70\% median Block Error Rate (BLER) during active browsing tests, driving a 36.6\% page-load failure rate. Conversely, the dense stadium Wi-Fi infrastructure delivers downlink throughput comparable to the best performing 5G Standalone (SA) deployment while providing better uplink and latency resilience, yielding the lowest game-day page-load failure rate (3.9\%) and bounding image upload latency degradation to just 2.1$\times$ relative to empty-stadium baselines. These insights proves that densification through localized Wi-Fi deployment is essential to absorb severe stadium traffic spikes.

\end{abstract}

\maketitle

\section{Introduction \& Background} \label{sec_introduction}

Modern stadiums present uniquely challenging wireless environments where crowd densities exceeding 50,000 active users generate massive localized capacity strain~\cite{ookla_superbowl}. This density is compounded by a structural shift toward symmetric traffic patterns, with uplink data accounting for over 30\% of total volume during major events~\cite{ericsson_mobility_events, nokia_mass_event_2020}. While operators address this via small cells, Distributed Antenna Systems (DAS), and mid-band/mmWave spectrum~\cite{xu2020operational5g, rochman2025comprehensive}, enterprise Wi-Fi 6E/7 deployments simultaneously leverage 1.2~GHz of unlicensed 6~GHz spectrum to offload carrier macro networks.

Unlike prior stadium measurement campaigns that focused strictly on raw signal propagation or incumbent interference~\cite{dogan2025evaluation}, characterizing actual user-facing performance under peak event loads remains critical. Extreme crowd density acutely amplifies the classic ``uplink gap''~\cite{khan2025mature, ghoshal2022indepth, chmieliauskas2025uplinkqoe}: while downlink infrastructure has matured, asymmetric resource allocation leaves the uplink fragile under load. Paradoxically, despite superior peak speeds, 5G can introduce severe uplink jitter and latency variance that degrades application-layer Quality of Experience (QoE) relative to stable LTE links~\cite{chmieliauskas2025uplinkqoe}.

To bridge this gap, this paper presents a data-driven analysis of user-facing QoE metrics at Notre Dame Stadium during sold-out football games attended by up to 77,622 spectators. To our knowledge, this work is the first to offer a comprehensive empirical comparison of multi-carrier 4G/5G cellular performance against 5/6~GHz Wi-Fi deployments within an active ultra-dense environment.
Our analysis provides three core contributions:

\noindent$\bullet$ \textbf{Large-Scale Empirical Characterization:} A user-centric evaluation across high-density cellular networks (up to 129 unique PCIs per operator) and a sprawling outdoor Wi-Fi deployment ($\sim$900 outdoor Access Points/APs) under massive stadium loads (77,622 spectators).

\noindent$\bullet$ \textbf{Quantifying Wi-Fi Latency Resilience:} Evidence that stadium Wi-Fi provides vastly superior uplink and latency resilience, yielding the lowest game-day page-load failure rate of 3.9\% and limiting media upload duration degradation to just 2.1$\times$ relative to empty-stadium baselines.

\noindent$\bullet$ \textbf{Cellular Uplink Gap Validation:} An empirical breakdown of an "uplink gap" under 5G Non-Standalone during congestion, driving a catastrophic 5,983~ms P90 browsing time-to-first-byte, a 46\% image upload failure rate, even with narrow low-band FDD channels demonstrating excellent channel quality during uplink throughput tests.

\begin{figure}[t]
  \centering
  \subfloat[Target sections \label{fig:target_sections}]{%
    \includegraphics[width=0.48\linewidth]{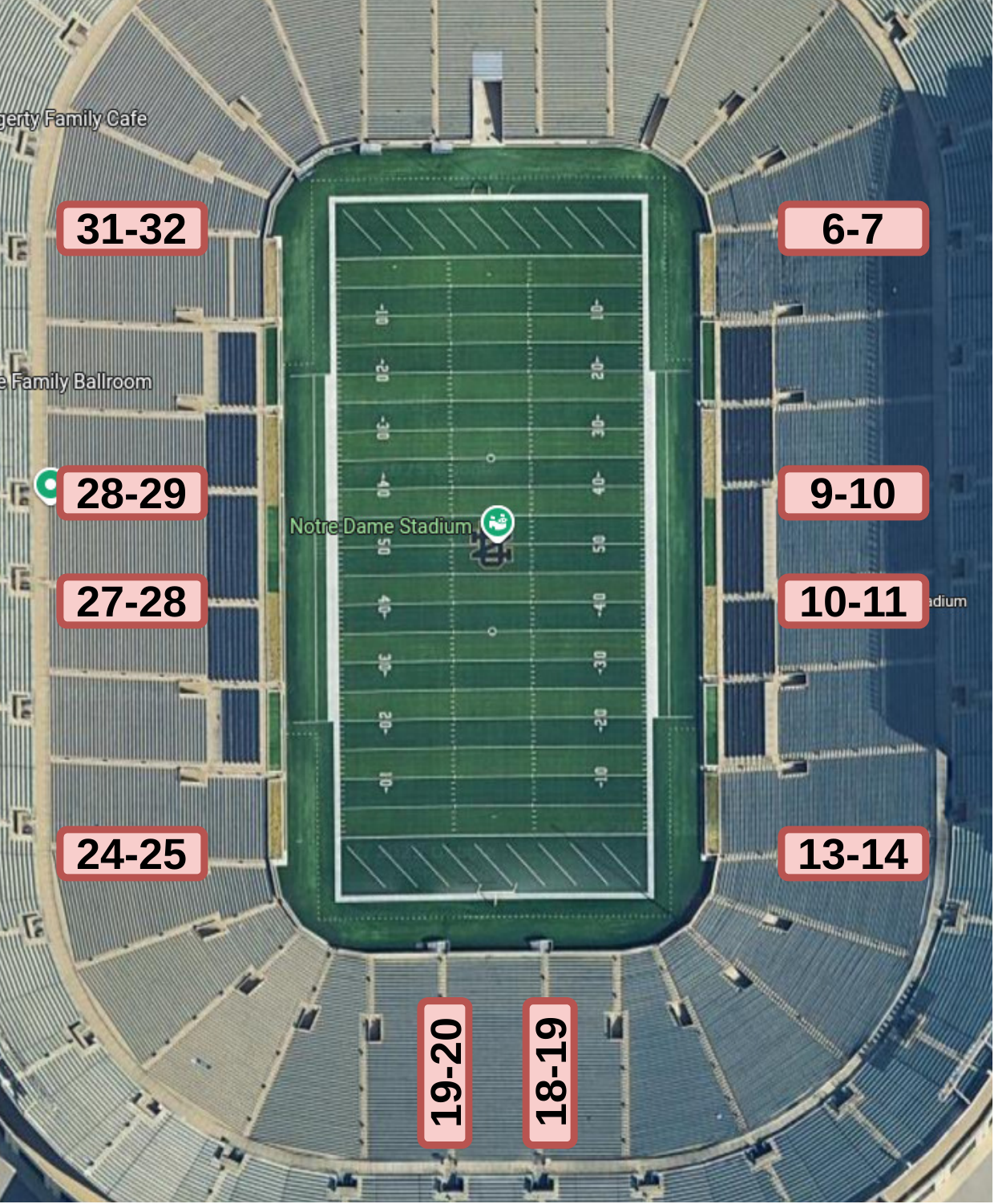}
  }
  \hfill
  \subfloat[Spots per section \label{fig:location_per_section}]{
    \includegraphics[width=0.48\linewidth]{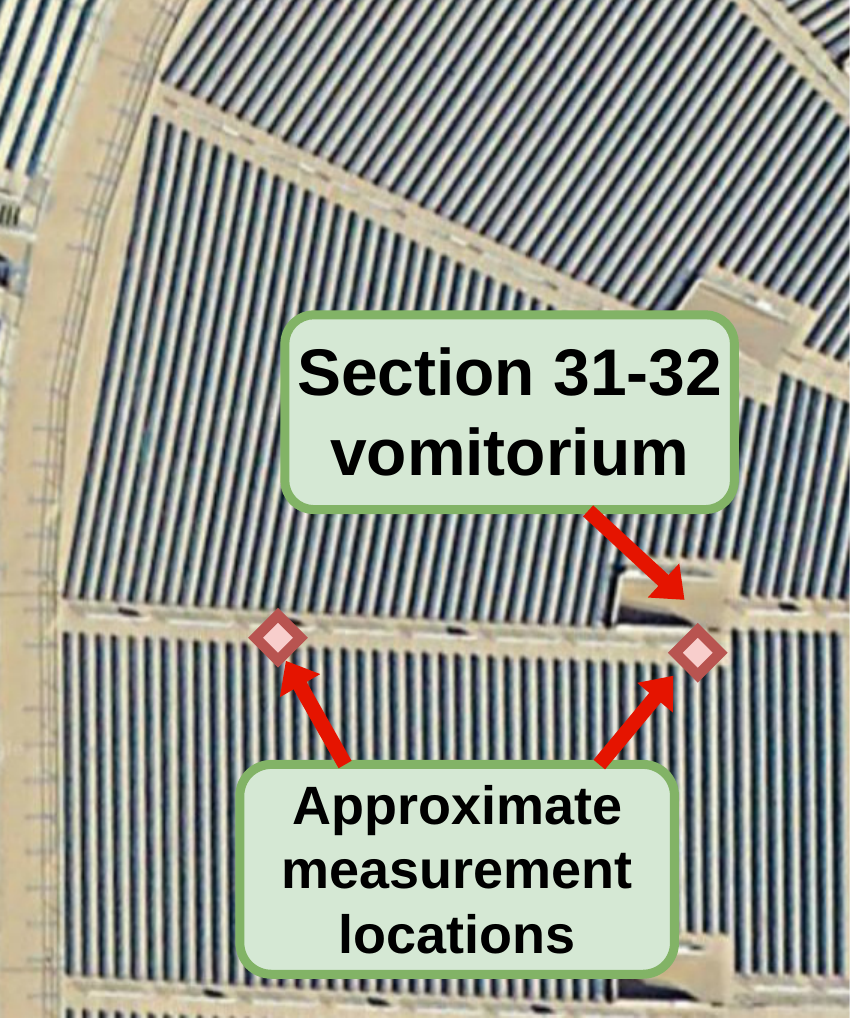}
  }
  
  \caption{Measurement spots at the stadium seating bowl.}
  \label{fig:measurement_target}
  \vspace{-1em}
\end{figure}



 
\section{Methodology and Tools} \label{sec:methodology}


\begin{table}
\centering
\caption{Summary of captured parameters.}
\label{tab:params}

\resizebox{\columnwidth}{!}{%
\begin{tabular}{|p{0.34\linewidth}|p{0.76\linewidth}|}
\hline
\textbf{Parameter} & \textbf{Description} \\
\hline\hline

\multicolumn{2}{|c|}{\textbf{\textit{Qualipoc \& SigCap: General parameters}}} \\ \hline
Latitude, Longitude & UE's geographic coordinates \\ \hline

\multicolumn{2}{|c|}{\textbf{\textit{Qualipoc: Radio parameters}}} \\ \hline
PCI & Physical Cell Identifier \\ \hline
eNB-ID & eNodeB identifiers, \ie 20-bit identifier for a specific 4G LTE cell tower or base station. \\ \hline
DL/UL ARFCN & Absolute Radio Frequency Channel Number, \ie center frequency \\ \hline
Bandwidth & Range of frequencies available for transmission [MHz] \\ \hline
RSRP & Reference Signal Received Power. For NR, RSRP is based on the Synchronization Signal (SS) block [dBm] \\ \hline
MCS & Modulation and coding scheme \\ \hline

\multicolumn{2}{|c|}{\textbf{\textit{Qualipoc: Application-based metrics}}} \\ \hline
DL/UL Throughput & Downlink/uplink throughput measured via Ookla Speedtest [Mbps]. \\ \hline
Idle Latency & Round-trip time (RTT) measured to the Ookla Speedtest server [ms]. \\ \hline
Browsing Duration & Delay from the initial HTTP URL request to the complete download of page resources [s]. \\ \hline
Time To First Byte (TTFB) & Delay from the initial HTTP request to the receipt of the first byte [s]. \\ \hline
First Second Throughput (FST) & The average data transfer rate measured in kilobytes during the first second of active download after connection establishment [KB/s]. \\ \hline
WhatsApp Sending Duration & Delay from initiating a message and 1~MB image transmission to receiving the delivery receipt [s]. \\ \hline
Instagram Posting Duration & Delay from initiating a 1~MB image upload to its successful completion [s]. \\ \hline 

BLER & Block error rate [\%] \\ \hline



\multicolumn{2}{|c|}{\textbf{\textit{SigCap: Wi-Fi parameters}}} \\ \hline
BSSID & Basic Service Set Identifier (unique identifier of a Wi-Fi AP) \\ \hline
Primary channel number & Primary channel associated with the BSSID [MHz] \\ \hline
RSSI & Received Signal Strength Indicator from the beacon signal [dBm] \\ \hline
Channel utilization & Fraction of time the 20~MHz primary channel is busy [0,1] \\ \hline

\end{tabular}
}%
\vspace{-1em}
\end{table}

Measurements spanned multiple sold-out football games (\textbf{game}) and corresponding vacant-stadium periods (\textbf{non-game}) to isolate the effects of network load. During the 2025 season, active throughput tests (Ookla Speedtest) were collected during two early-season games (09/13 and 09/27) and application metrics (\ie browsing, WhatsApp, Instagram) during two late-season games (11/08 and 11/22). Baseline \textbf{non-game} campaigns were conducted on 08/26/2025, 08/28/2025, and 02/26/2026. To the best of our knowledge, no physical infrastructure modifications has occurred between the measurement windows.

This work focuses exclusively on the outdoor stadium seating area (\textbf{bowl}). Due to physical constraints during events, data collection was restricted to the representative sections and measurement locations mapped in Figure~\ref{fig:measurement_target}. Measurements were gathered via six commercial UEs (five Samsung Galaxy S22+, one S24+) systematically cycling across stadium Wi-Fi and the three major U.S. carriers (AT\&T, T-Mobile, Verizon). No statistically significant baseline performance variance was observed across device chipsets.



For fine-grained logging, three UEs ran Rohde \& Schwarz QualiPoc to decode low-level Qualcomm modem signaling messages and extract layer-1/2 KPIs~\cite{qualipoc}. Moreover, QualiPoc is instrumental in automating all application testing (Ookla, browsing, WhatsApp, Instagram) and capturing QoE metrics discussed in this work. Concurrently, all UEs executed SigCap to record passive Wi-Fi beacon frames~\cite{dougan2025spectrum}. Table~\ref{tab:params} details the parameters used. In total, we collected 1,346,589 QualiPoc data samples and 1,889,707 SigCap Wi-Fi beacon samples.

\begin{table}[t]
    \centering
    \caption{Summary of LTE and 5G NR carrier configurations.
        }
    \label{tab:carrier_config}
    \resizebox{\columnwidth}{!}{%
    \begin{tabular}{|c|c|c|c|c|c|c|c|}
        \hline
        \multirow{2}{*}{\textbf{Operator}} & \multirow{2}{*}{\textbf{Band}} & \multirow{2}{*}{\textbf{Duplex}} & \multicolumn{2}{c|}{\textbf{Freq. (MHz)\(^\ddagger\)}} & \textbf{BW} & \multicolumn{2}{c|}{\textbf{\# PCI}} \\ \cline{4-5}\cline{7-8}
        & & & \textbf{UL} & \textbf{DL} & \textbf{(MHz)} & \textbf{GD\(^\dagger\)} & \textbf{NG\(^\dagger\)} \\
        \hline
        \hline
        
         \multicolumn{8}{|c|}{\textbf{NR Bands}} \\
        \hline
        
        \multirow{1}{*}{AT\&T} & n5 & FDD & 829 & 874 & 10 & 25& 30\\
        \hline
        
        \multirow{3}{*}{T-Mobile}
        & n25 & FDD & 1894 & 1974 & 20 & 23& 3\\ \cline{2-8}
        & n71 & FDD & 677 & 631 & 20& 1& 0\\ \cline{2-8}
        & n41 & TDD & \multicolumn{2}{c|}{2507, 2607} & 100, 90 & 38& 29\\
        \hline
        
        \multirow{1}{*}{Verizon} 
        & n77 & TDD & \multicolumn{2}{c|}{3730, 3809} & 100 & 11& 7\\
        \hline
        \hline
        
         \multicolumn{8}{|c|}{\textbf{LTE Bands}} \\
        \hline
        
        \multirow{5}{*}{AT\&T} 
        & b2 & FDD & 1860 & 1940 & 20 & 32& 35\\ \cline{2-8}
        & b4 & FDD & 1735 & 2135 & 10& 10& 0\\ \cline{2-8}
        & b12 & FDD & 709 & 739 & 10 & 33& 36\\ \cline{2-8}
        & b30 & FDD & 2310 & 2355 & 10 & 27& 32\\ \cline{2-8}
        & b66 & FDD & 1735 & 2135 & 10& 31& 33\\
        \hline
        
        \multirow{2}{*}{T-Mobile}
        & b2 & FDD & 1875 & 1955 & 10& 23 & 1\\ \cline{2-8}
        & b66 & FDD & 1748 & 2148 & 15 & 23& 2\\ 
        \hline
        
        \multirow{6}{*}{Verizon} 
        & b2 & FDD & 1883 & 1963 & 5 & 22& 26\\ \cline{2-8}
        & b4 & FDD & 1720 & 2120 & 20& 14& 2\\ \cline{2-8}
        & b5 & FDD & 840 & 885 & 10 & 24& 17\\ \cline{2-8}
        & b13 & FDD & 782 & 751 & 10 & 24& 24\\ \cline{2-8}
        & b48 & TDD& \multicolumn{2}{c|}{3600} & 20& 1& 0\\ \cline{2-8}
        & b66 & FDD & 1720 & 2120 & 20 & 30& 26\\
        \hline
        
    \end{tabular}
    }
    \(^\dagger\) \textbf{NG} and \textbf{GD} refer to \textbf{non-game} and \textbf{game} measurements, respectively. \\
    \(^\ddagger\) Frequencies are rounded to the nearest 1 MHz. 

    \vspace{-1em}
\end{table}

\section{Deployment Overview}

\subsection{Cellular Deployment}

Our measurements across 6 games in the 2025 football season a high density of Physical Cell Identifiers (PCIs), reaching up to 129 unique 4G and 5G PCIs per operator. Because a third-party provider manages the stadium infrastructure to host the three major cellular carriers, this high PCI count reflects a neutral-host deployment utilizing aggressive sectorization via small cells/DAS to mitigate extreme crowd density. In this work, we constrain our analysis to the specific four days of football and three non-game days dedicated to application-based metrics, as described in Section \ref{sec:methodology}. Table~\ref{tab:carrier_config} details the specific 4G and 5G carriers extracted from this filtered dataset, along with the number of unique observed PCIs. Overall, PCI assignment to the DAS radio unit changes between game and non-game measurements. For many bands, especially on T-Mobile and Verizon, fewer PCIs are observed during non-game periods. AT\&T LTE bands show a different pattern, with PCI counts often higher when the stadium is empty.

\subsection{Wi-Fi Deployment}

The enterprise Wi-Fi infrastructure in the stadium utilizes an HPE Aruba system consisting of Wi-Fi 6E Standard Power (SP) and Low-Power Indoor (LPI) access points (APs); specifically, the AP-634 and AP-635 models, respectively. The deployment includes $\sim$900 SP APs installed along the railings between seating sections of the outdoor bowl, alongside $\sim$400 LPI APs distributed across indoor concourses, concession areas, and adjacent VIP premium seating structures. By focusing exclusively on the outdoor bowl, this analysis isolates the performance of outdoor SP APs. Each AP is configured to broadcast beacons in both the 5~GHz and 6~GHz bands, with data channel width of 20~MHz and 80~MHz, respectively.

\section{Results \& Discussions} \label{sec_results}

\begin{table}[t]
\centering
\small
\caption{RAN technology deployment summary: Game vs. non-game.
}
\label{tab:RAN_Summary}
\resizebox{.9\columnwidth}{!}{%
\begin{tabular}{|l|l|c|c|}
\hline
\textbf{Operator} & \textbf{Technology} & \textbf{Game (\%)} & \textbf{Non-game (\%)} \\ \hline \hline
\multirow{2}{*}{AT\&T} & 5G EN-DC & 55.65& 57.41\\
                       & LTE      & 44.35& 42.59\\ \hline
\multirow{3}{*}{T-Mobile} & 5G EN-DC& 49.64& 0.00\\
                           & LTE      & 8.13& 3.93\\
                           & 5G SA      & 42.23& 96.07\\ \hline
\multirow{2}{*}{Verizon} & 5G EN-DC & 64.54& 29.13\\
                         & LTE      & 35.46& 70.87\\ \hline

\end{tabular}
}
\vspace{-1em}
\end{table}

\subsection{Deployment Analysis}

\subsubsection{Cellular Deployments}

Table~\ref{tab:RAN_Summary} reveals distinct architectural strategies between carriers across measurement periods. T-Mobile primarily operates in 5G Standalone (SA) mode via its n41 and n25 bands (Table~\ref{tab:carrier_config}), accounting for 96.07\% of its non-game baseline. Conversely, AT\&T and Verizon rely on 5G Non-Standalone (NSA) mode via E-UTRAN New Radio Dual Connectivity (EN-DC), where an LTE connection serves as the anchor layer. Under active game-day crowd loads, behavioral divergence becomes stark: AT\&T's EN-DC utilization remains flat at $\sim$56\%, whereas Verizon's EN-DC usage surges from 29.13\% to 64.54\%, signaling a tactical shift toward its capacity-focused mid-band n77 spectrum to absorb the extreme traffic load.

As we observe high number of PCIs which suggests a small-cell DAS architecture, we further identified consecutive 4G eNodeB identifiers unique to the stadium. These identifiers are further cross-correlated with external data to confirm  its exclusivity to the stadium deployment.
Specifically, we observed unique consecutive eNB-IDs with prefixes ``7260'' (11 IDs) for AT\&T, ``2087'' (16 IDs) for Verizon, and ``6264'' (6 IDs) for T-Mobile. These served all low- and mid-band 4G channels listed in Table~\ref{tab:carrier_config}.
Although no direct metric indicated a 5G small-cell DAS deployment, we posit its presence due to the observation of numerous unique PCIs, similar to the LTE results.

\begin{figure}[t]
    \centering
    \begin{subfigure}[t]{0.9\linewidth}
        \centering
        \includegraphics[width=\linewidth]{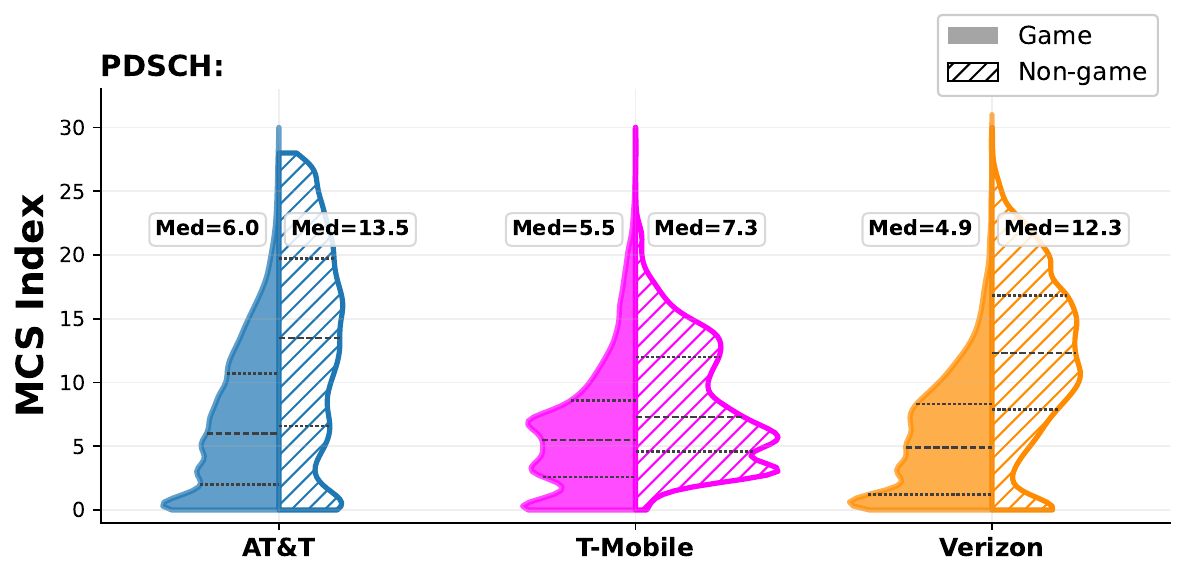}
        \vspace{-1.5em}
        \caption{}
        \label{subfig:MCS-violin-dl}
    \end{subfigure}
    \begin{subfigure}[t]{0.9\linewidth}
        \centering
        \includegraphics[width=\linewidth]{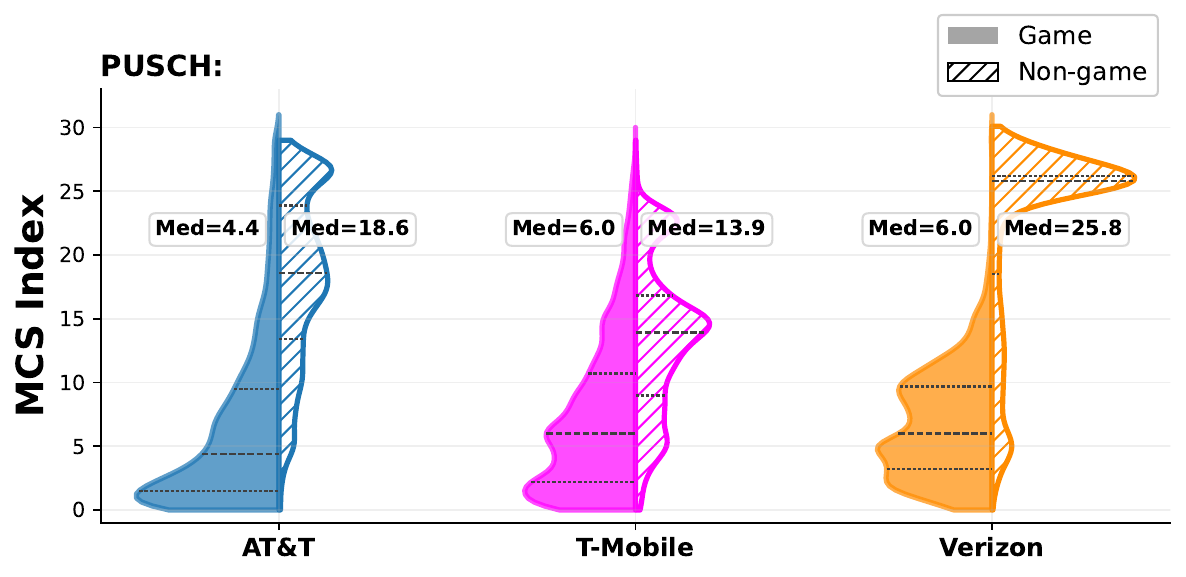}
        \vspace{-1.5em}
        \caption{}
        \label{subfig:MCS-violin-ul}
    \end{subfigure}
    \caption{Comparison of (a) downlink and (b) uplink MCS index.}
    \label{fig:MCS-violin}
\end{figure}

The 4G DAS analysis is further corroborated by the robust signal strength profiles, maintaining a stable median LTE RSRP of -89~dBm in both low- and mid-band channels during the games. Similarly, a median of -83~dBm is observed in low- and mid-bands during non-game periods. However, median NR RSRP profiles during the games reveal spectrum-dependent propagation characteristics: AT\&T's low-band n5 achieved a strong median SS-RSRP of -75~dBm during games, T-Mobile's mid-band channels achieved a moderate median of -92~dBm, and Verizon's higher-frequency 3.7~GHz n77 registered a weaker -106~dBm. Ultimately, localized crowd density severely degrades user-perceived channel quality. As illustrated by the joint 4G/5G MCS distributions (Fig.~\ref{fig:MCS-violin}), heavy game-day traffic triggers aggressive downward link adaptation. Median downlink MCS configurations collapse for AT\&T (dropping from 13.5 to 6.0) and Verizon (from 12.3 to 4.9), whereas T-Mobile maintains a lower MCS reduction from 7.3 to 5.3. The uplink channel experiences even greater degradation with steep declines of median MCS: AT\&T drops from 18.6 to 4.4, T-Mobile from 13.9 to 6.0, and Verizon from 25.8 down to 6.0.

\subsubsection{Wi-Fi Deployment Analysis}
\label{sec:wifi_deployment}

\begin{figure}[t]
    \centering
    \includegraphics[width=1\linewidth]{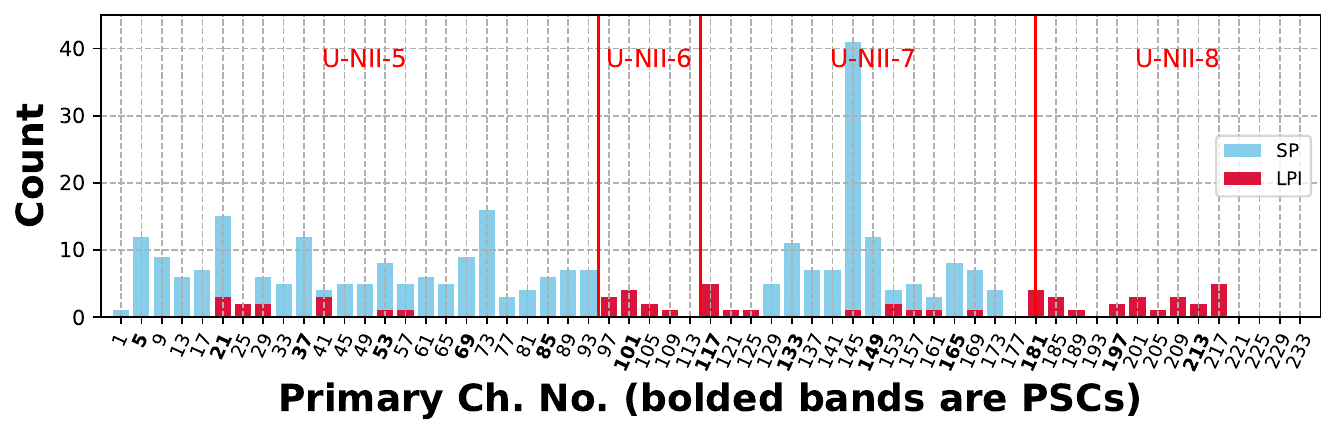}
    \vspace{-2em}
    \caption{Count of unique BSSIDs on 2026 non-game campaign.}
    \label{fig:wifi_bssid_2026}
    \vspace{-6pt}
\end{figure}

Our 2025 game and non-game measurements within the outdoor bowl revealed a uniform allocation of SP BSSIDs across Preferred Scanning Channels (PSCs) in the U-NII-5 and U-NII-7 bands. Conversely, LPI deployments utilized PSCs across the entire 6~GHz spectrum, with higher concentrations in the U-NII-6 and U-NII-8 bands. This channel assignment aligns directly with our findings from the 2024 football season~\cite{dogan2025evaluation}, thus we omit this analysis. However, as illustrated in Fig.~\ref{fig:wifi_bssid_2026}, our 2026 non-game measurements reveal a deployment strategy that is no longer constrained by strict PSC assignments. One possible explanation for the use of non-PSC allocations in the enterprise network is the potential for aggregate co-channel interference on the PSC in dense and shared environments, a vulnerability highlighted in our previous study~\cite{dogan2025evaluation}.


\begin{figure}[t]
    \centering
    \begin{subfigure}[t]{0.37\linewidth}
        \centering
        \includegraphics[width=\linewidth]{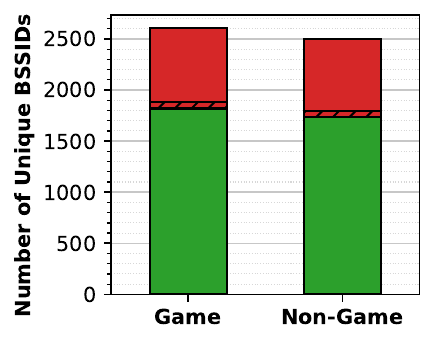}
        \vspace{-1.5em}
        \caption{Count of Unique BSSIDs}
        \label{subfig:unique_bssid}
    \end{subfigure}
    \begin{subfigure}[t]{0.61\linewidth}
        \centering
        \includegraphics[width=\linewidth]{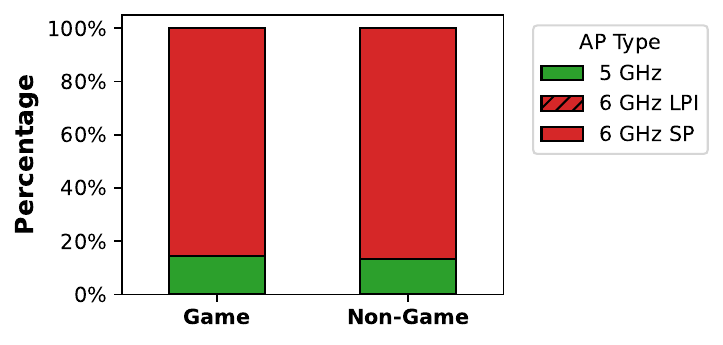}
        \vspace{-1.5em}
        \caption{Connection Ratio}
        \label{subfig:conn_ratio}
    \end{subfigure} \\
    
    \begin{subfigure}{0.8\linewidth}
        \centering
        \includegraphics[width=\linewidth]{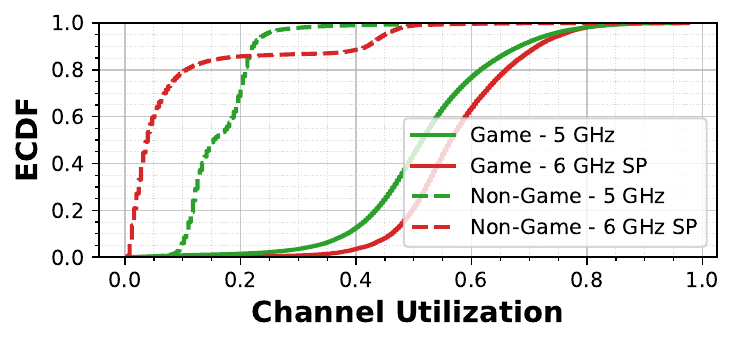}
        \vspace{-1.5em}
        \caption{Channel Utilization}
        \label{subfig:ch_util}
    \end{subfigure}
    \vspace{-.5em}
    \caption{Comparison of Wi-Fi band utilization across game and non-game periods.}
    \label{fig:wifi_band_comp}
    \vspace{-1em}
\end{figure}



\begin{figure*}[t]
    \centering
    \begin{subfigure}[t]{0.325\linewidth}
        \centering
        \includegraphics[width=\linewidth]{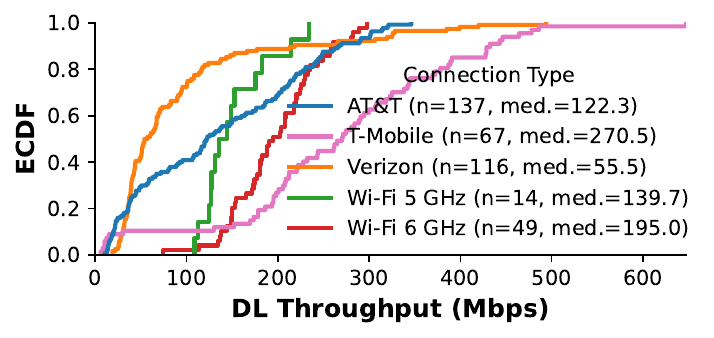}
        \vspace{-1.5em}
        \caption{}
        \label{subfig:ookla_pregame_dl}
    \end{subfigure}
    \begin{subfigure}[t]{0.325\linewidth}
        \centering
        \includegraphics[width=\linewidth]{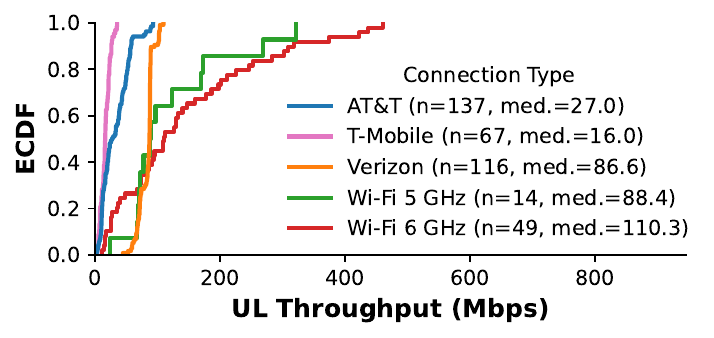}
        \vspace{-1.5em}
        \caption{}
        \label{subfig:ookla_pregame_ul}
    \end{subfigure}
    \begin{subfigure}[t]{0.325\linewidth}
        \centering
        \includegraphics[width=\linewidth]{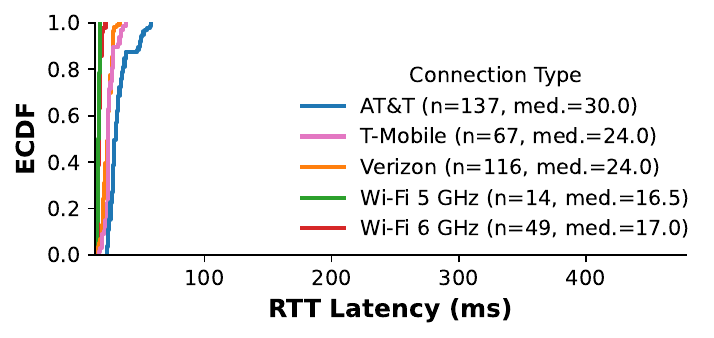}
        \vspace{-1.5em}
        \caption{}
        \label{subfig:ookla_pregame_lat}
    \end{subfigure}
    \vspace{-1em}
    \caption{Comparison of Ookla Speedtest metrics during non-game periods.}
    \label{fig:ookla_pregame}
\end{figure*}

\begin{figure*}[t]
    \centering
    \begin{subfigure}[t]{0.325\linewidth}
        \centering
        \includegraphics[width=\linewidth]{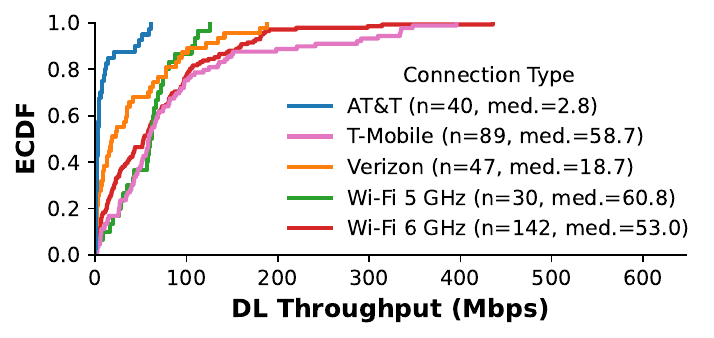}
        \vspace{-1.5em}
        \caption{}
        \label{subfig:ookla_game_dl}
    \end{subfigure}
    \begin{subfigure}[t]{0.325\linewidth}
        \centering
        \includegraphics[width=\linewidth]{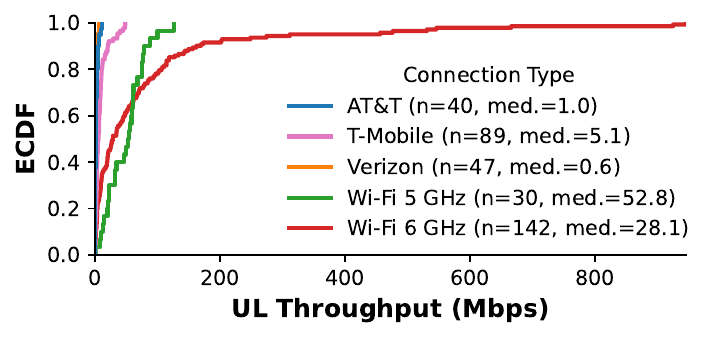}
        \vspace{-1.5em}
        \caption{}
        \label{subfig:ookla_game_ul}
    \end{subfigure}
    \begin{subfigure}[t]{0.325\linewidth}
        \centering
        \includegraphics[width=\linewidth]{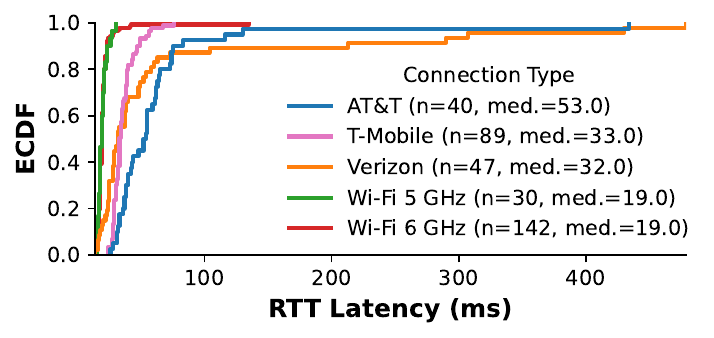}
        \vspace{-1.5em}
        \caption{}
        \label{subfig:ookla_game_lat}
    \end{subfigure}
    \vspace{-1em}
    \caption{Comparison of Ookla Speedtest metrics during game periods.}
    \label{fig:ookla_game}
\end{figure*}

Across all campaigns, devices observed a stable footprint of broadcasting BSSIDs across the 5~GHz, 6~GHz SP, and 6~GHz LPI bands (Fig.~\ref{subfig:unique_bssid}). The count of detected 5~GHz BSSIDs consistently outpaces 6~GHz bands due to more efficient client scanning over the narrower 5~GHz space, particularly because discovery was not restricted to 6~GHz PSCs (validated by non-PSC detections in Fig.~\ref{fig:wifi_bssid_2026}).

Conversely, Fig.~\ref{subfig:conn_ratio} highlights a clear client preference for 6~GHz SP APs across both periods, driven by superior signal quality. Connected 6~GHz SP sessions maintained strong median RSSIs of -52~dBm (game) and -56~dBm (non-game), outperforming 5~GHz connections by 5~dB and 10~dB, respectively. Conversely, 6~GHz LPI connectivity was minimal due to its indoor-only concourse deployment. This association bias correlates to the channel utilization extracted from BSS Load Elements of connected and neighboring beacons (Fig.~\ref{subfig:ch_util}). While 6~GHz utilization exceeds 5~GHz levels during games, it falls below 5~GHz baselines during empty periods, indicating a higher adoption of 6~GHz-capable devices that concentrates games' traffic onto the newer spectrum band.

\subsection{Comparison of Ookla Speedtest Metrics}
\label{sec:ookla_st}

During our analysis, we identified a performance trade-off inherent to our measurement methodology. To capture the dynamic state of the Wi-Fi network (\eg channel utilization) concurrently with Ookla Speedtests, SigCap was configured to aggressively scan the Wi-Fi bands every 5 seconds. However, this active scanning introduced significant radio-interface overhead, resulting in throughput throttling. Controlled non-game baseline experiments conducted on 02/26/2026 revealed that this active scanning overhead reduced Wi-Fi throughput by a median of 52.91\% on the downlink and 47.62\% on the uplink. To compensate for this systematic measurement artifact, we applied empirical scaling coefficients of 1.89 and 2.10 to the Wi-Fi downlink and uplink throughput datasets, respectively. Notably, latency metrics, which were recorded in an ``idle'' state prior to active throughput testing, remained unaffected by this scanning overhead.

Figs.~\ref{fig:ookla_pregame} and \ref{fig:ookla_game} compare Ookla Speedtest DL throughput, UL throughput, and RTT latency metrics, which were presumably evaluated against the closest optimal local server. As expected, throughput decreases and latency increases significantly during games compared to non-game baselines. The sample sizes across connection types vary due to minor fluctuations in sampling intervals, day-to-day variations in the specific phone models and connection modes monitored, and the exclusion of failed sessions. Accordingly, the illustrated distributions represent successful tests exclusively.

Fig.~\ref{subfig:ookla_pregame_dl} illustrates the baseline peak DL throughput achieved by T-Mobile with a median of 270.5~Mbps during vacant, non-game periods. Additionally, Wi-Fi 6~GHz connections exhibit higher non-game DL throughput compared to the 5~GHz band; this behavior is expected given the operational differences in channel bandwidth (20~MHz vs. 80~MHz) configured in the stadium. However, as shown in Fig.~\ref{subfig:ookla_game_dl}, this performance gap narrows significantly during games. Under these high channel utilization condition demonstrated in Fig.~\ref{subfig:ch_util}, we observe highly comparable peak DL throughput performance across Wi-Fi 5~GHz, Wi-Fi 6~GHz, and T-Mobile cellular connections, with a median DL throughput between 53 and 60.8~Mbps.
Figs.~\ref{subfig:ookla_pregame_ul} and \ref{subfig:ookla_game_ul} shows the best UL throughput performance achieved by Wi-Fi connection, both during the game and non-game periods. In particular, Wi-Fi 5~GHz slightly edges its 6~GHz counterparts with a notable median UL throughput of 52.8~Mbps vs. 28.1~Mbps.
Similarly, Figs.~\ref{subfig:ookla_pregame_lat} \& \ref{subfig:ookla_game_lat} also shows the Wi-Fi connection outclassed cellulars in latency, with the lowest median of 16.5~ms achieved by 5~GHz during non-game, and the lowest median of 19~ms tied between 5 and 6~GHz during the games.

Comparing the cellular operators, AT\&T experiences the most severe performance degradation in both throughput and latency during games relative to non-game baselines. While the network predominantly utilizes 5G-NSA mode during testing, it relies strictly on the narrow 10~MHz n5 channel, lacking access to wider 3.5~GHz mid-band (C-band) coverage within the stadium. Consequently, AT\&T faces a severe capacity deficit under dense user crowds. This stands in stark contrast to T-Mobile and Verizon, which successfully mitigate game-day congestion by leveraging 190~MHz and 200~MHz of mid-band spectrum, respectively.

During game day, T-Mobile outperforms both Verizon and AT\&T. This advantage stems from a denser PCI deployment (Table \ref{tab:carrier_config}) and the favorable propagation characteristics of T-Mobile’s 2.5 GHz n41 band compared to Verizon’s higher-frequency 3.7 GHz n77 band. This propagation difference is directly reflected in the MCS selections. Specifically, T-Mobile maintained a median DL MCS of 11 and a UL MCS of 5 on its n41 band, whereas Verizon's n77 band managed only a median DL MCS of 8 and a UL MCS of 3 during Game Day.

Interestingly, the non-game uplink data presents a reversed trend, with AT\&T and Verizon delivering higher uplink throughput than T-Mobile. A per-band analysis reveals that during the non-game period, the vast majority of uplink traffic---a median of over 80\% for both AT\&T and Verizon---was routed over low-frequency FDD LTE bands. These LTE bands sustained excellent channel quality, with both operators recording at least one LTE band with a median UL MCS above 22. By comparison, Verizon's mid-band n77 and T-Mobile's mid-band n41 achieved a non-game UL MCS of only 7 and 11, respectively which highlights the severe pathloss faced in these higher bands. AT\&T and Verizon maintained denser active LTE deployments during the pregame window (Table \ref{tab:carrier_config}), whereas T-Mobile's LTE infrastructure appeared to be absent during the non-game period and fully activated only during game day, leaving T-Mobile with the weakest non-game uplink performance. This behavior aligns with a trend observed throughout our measurements: most of the uplink load is routed through low-frequency FDD bands, while downlink throughput is dominated by wide, high-frequency TDD bands. This is explored in detail in \cite{shuvo2026comprehensive}.


\subsection{Comparison of Browsing Metrics}
\label{sec:browsing_test}

\begin{figure}[t]
    \centering
    \includegraphics[width=1\linewidth]{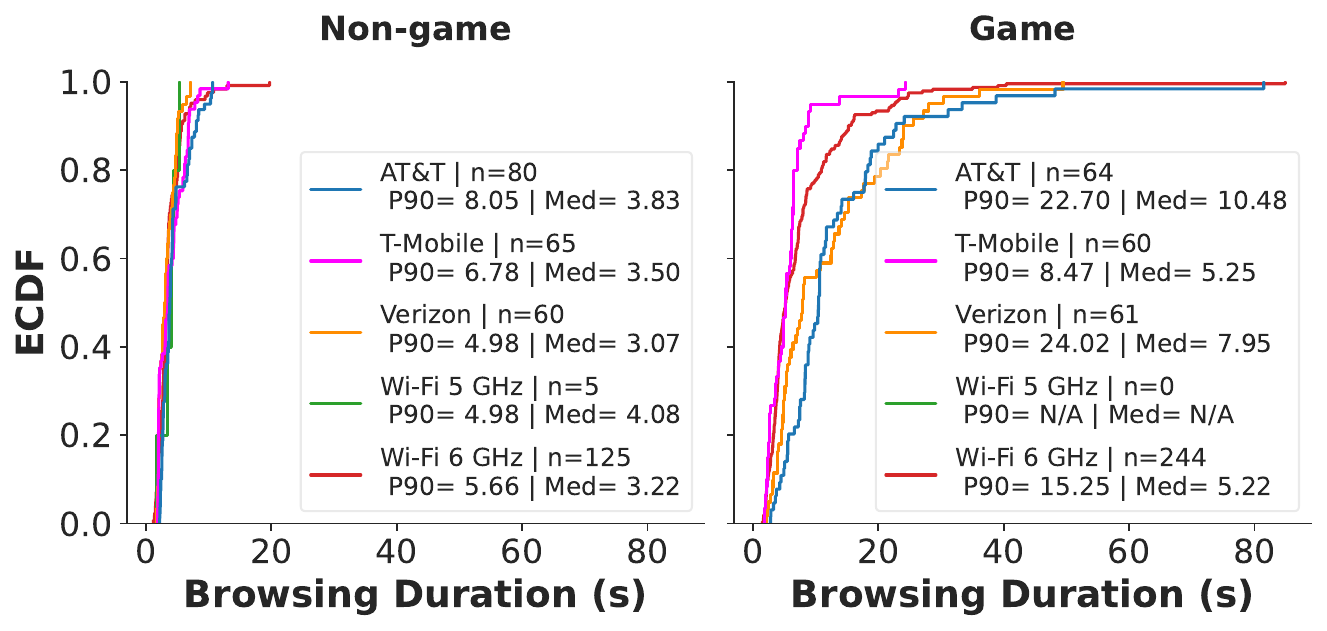}
    \caption{ECDF of user-perceived browsing duration, game-day versus non-game baseline.}
    \label{fig:ecdf-browsing-duration}
\end{figure}

 \begin{table}[t]
\centering
\caption{Browsing test performance comparison, showing sample size counts (Completed / Total), time to first byte (TTFB) and first-second throughput (FST).}
\label{tab:TTFB_FST_Comparison}
\resizebox{\columnwidth}{!}{%
\begin{tabular}{|l|cc|cc|cc|cc|cc|}
\hline
\multirow{3}{*}{\textbf{Operator}} & \multicolumn{2}{c|}{\textbf{Sample Counts}} & \multicolumn{4}{c|}{\textbf{TTFB (ms)}} & \multicolumn{4}{c|}{\textbf{FST (KB/s)}} \\ \cline{2-11} 
 & \multicolumn{2}{c|}{\textbf{(Completed / Total)}} & \multicolumn{2}{c|}{\textbf{Game}} & \multicolumn{2}{c|}{\textbf{Non-game}} & \multicolumn{2}{c|}{\textbf{Game}} & \multicolumn{2}{c|}{\textbf{Non-game}} \\ \cline{2-11} 
 & \textbf{Game} & \textbf{Non-game} & \textbf{Med} & \textbf{P90} & \textbf{Med} & \textbf{P90} & \textbf{Med} & \textbf{P10} & \textbf{Med} & \textbf{P10} \\ \hline \hline
AT\&T        & 64 / 101 & 80 / 80   & 691 & 2916 & 611 & 836 & 298 & 121 & 1179 & 482 \\ \hline
T-Mobile     & 60 / 80  & 65 / 65   & 504 & 825  & 454 & 781 & 687 & 379 & 856  & 406 \\ \hline
Verizon      & 61 / 81  & 60 / 60   & 750 & 5983 & 349 & 488 & 460 & 115 & 1261 & 705 \\ \hline
Wi-Fi 5 GHz  & 0 / 0    & 5 / 5     & -   & -    & 453 & 519 & -   & -   & 1018 & 874 \\ \hline
Wi-Fi 6 GHz  & 244 / 254& 125 / 125 & 573 & 1458 & 447 & 734 & 743 & 213 & 1301 & 449 \\ \hline
\end{tabular}
}
\end{table}

We evaluate browsing QoE using browsing duration (total latency from request to complete page load)
in Fig.~\ref{fig:ecdf-browsing-duration}, as well as network-layer bottlenecks via time to first byte (TTFB) and initial burst capacities via first-second throughput (FST) in Table~\ref{tab:TTFB_FST_Comparison}. The tests span over the following sites: \url{instagram.com}, \url{tiktok.com}, \url{facebook.com}, \url{ncaa.com}, and \url{sports.yahoo.com}. We also note that connections taking longer than 90 seconds are flagged as failed. Because failed tests do not report a browsing duration and selectively record TTFB, we filter the metrics to completed tests to ensure a fair cross-carrier comparison.

Shown in Table~\ref{tab:TTFB_FST_Comparison}, the Wi-Fi 6~GHz deployment anchored the venue with the lowest overall session failure rate at 3.9\%, while all three cellular networks clustered around a much higher failure tier, with T-Mobile at 25.0\%, Verizon at 24.7\%, and AT\&T at 36.6\%. Additionally, Fig.~\ref{fig:ecdf-browsing-duration} illustrates significant tail performance degradation during active game windows from its non-game baseline. By looking at the 90th percentile (P90) browsing duration, T-Mobile's proved to be the most stable, experiencing a tight 1.3$\times$ latency inflation. In comparison, Wi-Fi 6~GHz and AT\&T degraded by 2.7$\times$ and 2.8$\times$, respectively, while Verizon exhibited the most severe self-degradation, with its P90 tail exploding to 4.8$\times$ its baseline value. Due to automated band-steering under load, no gameday samples were captured on Wi-Fi 5~GHz, prompting its omission from game comparison.

\begin{figure*}[t]
    \centering
    \begin{subfigure}[t]{0.325\linewidth}
        \centering
        \includegraphics[width=\linewidth]{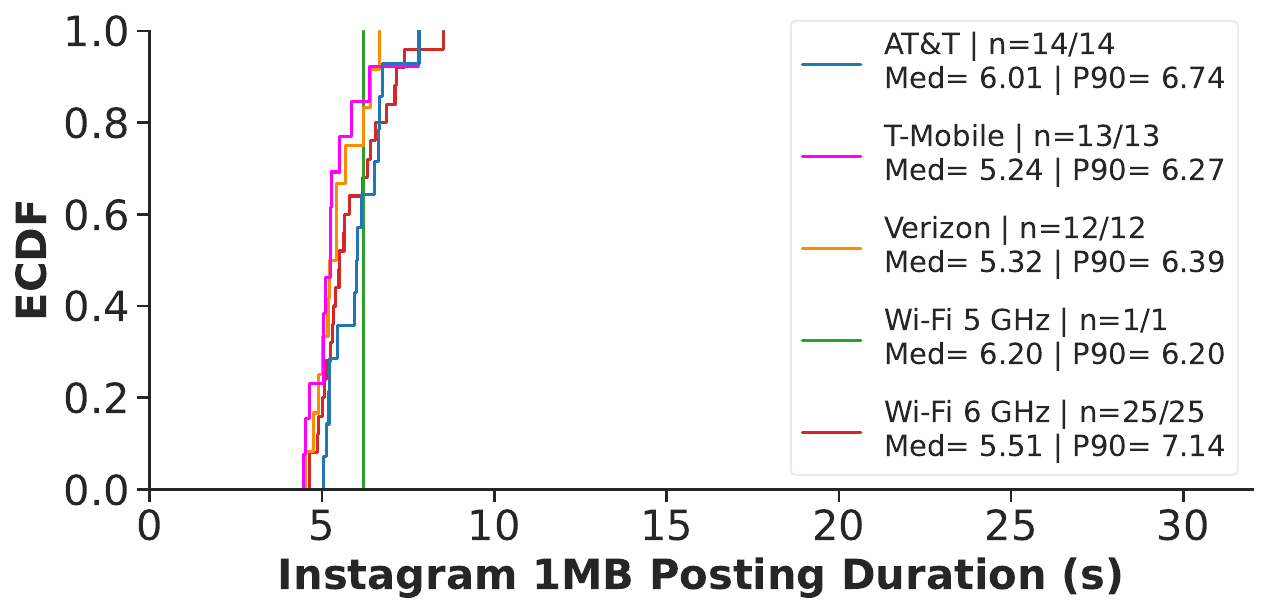}
        \vspace{-1.5em}
        \caption{Instagram 1~MB Post}
        \label{subfig:igwa_pregame_posting}
    \end{subfigure}
    \begin{subfigure}[t]{0.325\linewidth}
        \centering
        \includegraphics[width=\linewidth]{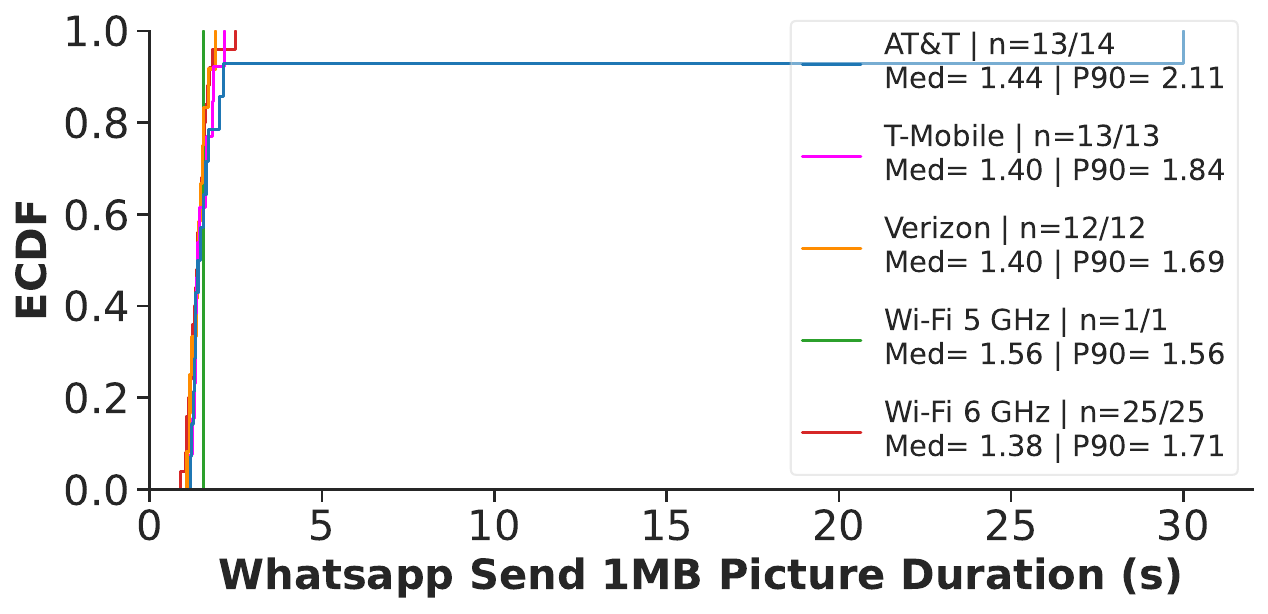}
        \vspace{-1.5em}
        \caption{WhatsApp 1~MB Image}
        \label{subfig:igwa_pregame_picture}
    \end{subfigure}
    \begin{subfigure}[t]{0.325\linewidth}
        \centering
        \includegraphics[width=\linewidth]{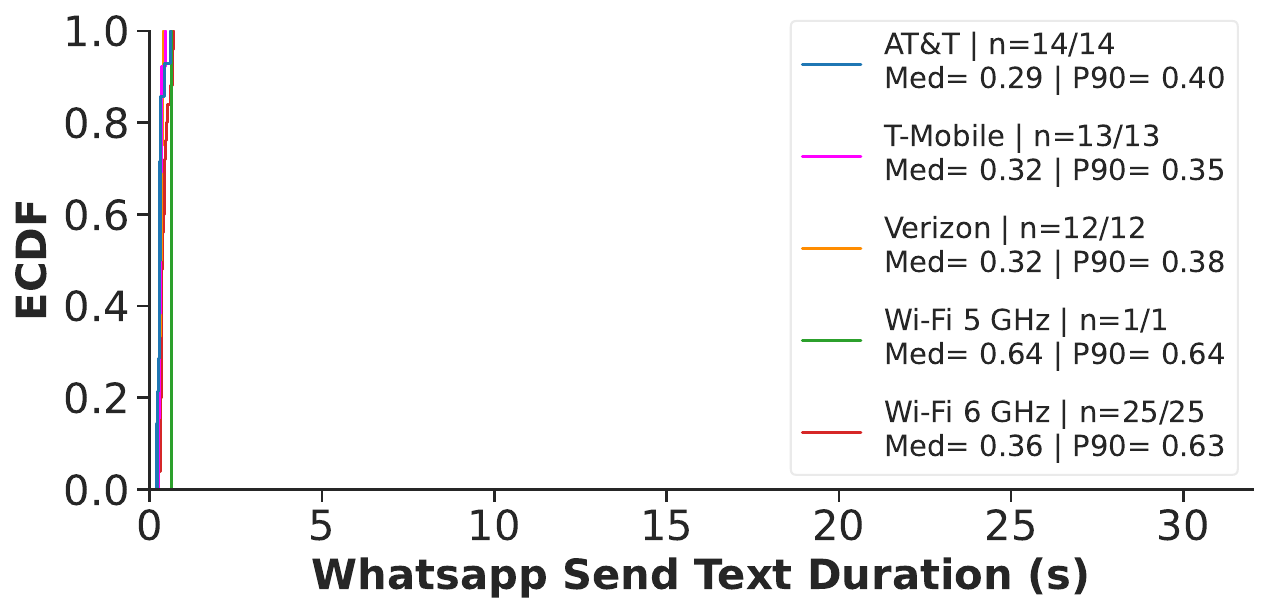}
        \vspace{-1.5em}
        \caption{WhatsApp Text}
        \label{subfig:igwa_pregame_text}
    \end{subfigure}
    \vspace{-.5em}
    \caption{WhatsApp Sending and Instagram Posting duration ECDF during non-game periods. Failed tests are imputed at a 30.00~s timeout ceiling. \textit{The size of the completed tests ($X$) out of all tests ($Y$) here is written as $n= X/Y$.}}
    \label{fig:igwa_pregame}
\end{figure*}

\begin{figure*}[t]
    \centering
    \begin{subfigure}[t]{0.325\linewidth}
        \centering
        \includegraphics[width=\linewidth]{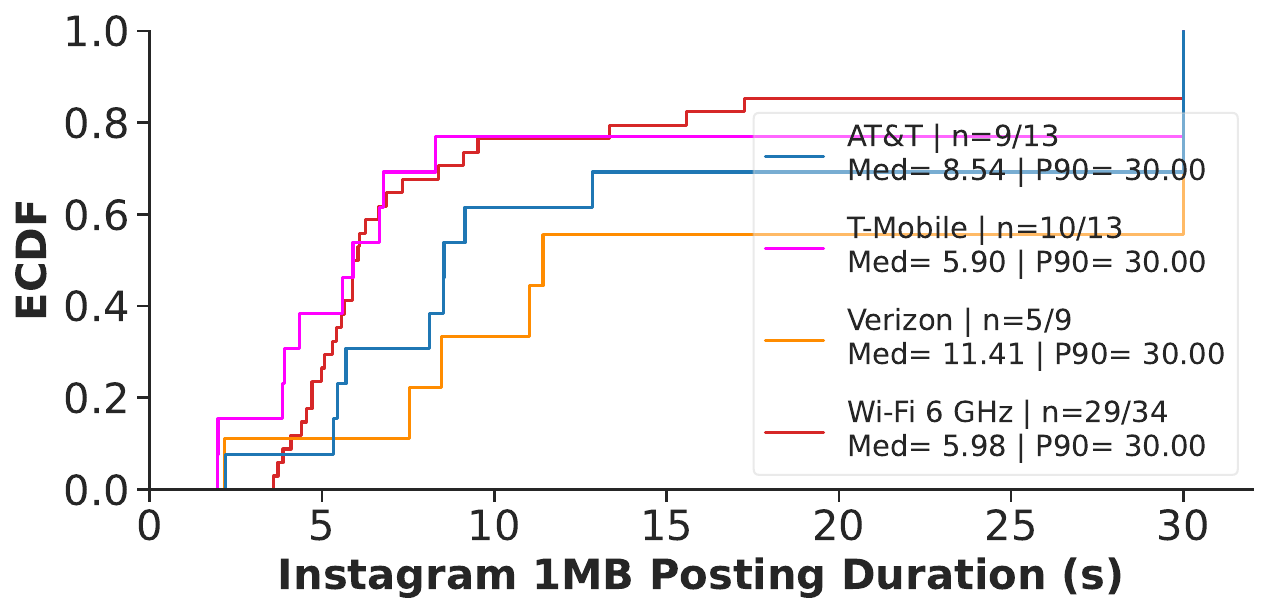}
        \vspace{-1.5em}
        \caption{Instagram 1~MB Post}
        \label{subfig:igwa_game_posting}
    \end{subfigure}
    \begin{subfigure}[t]{0.325\linewidth}
        \centering
        \includegraphics[width=\linewidth]{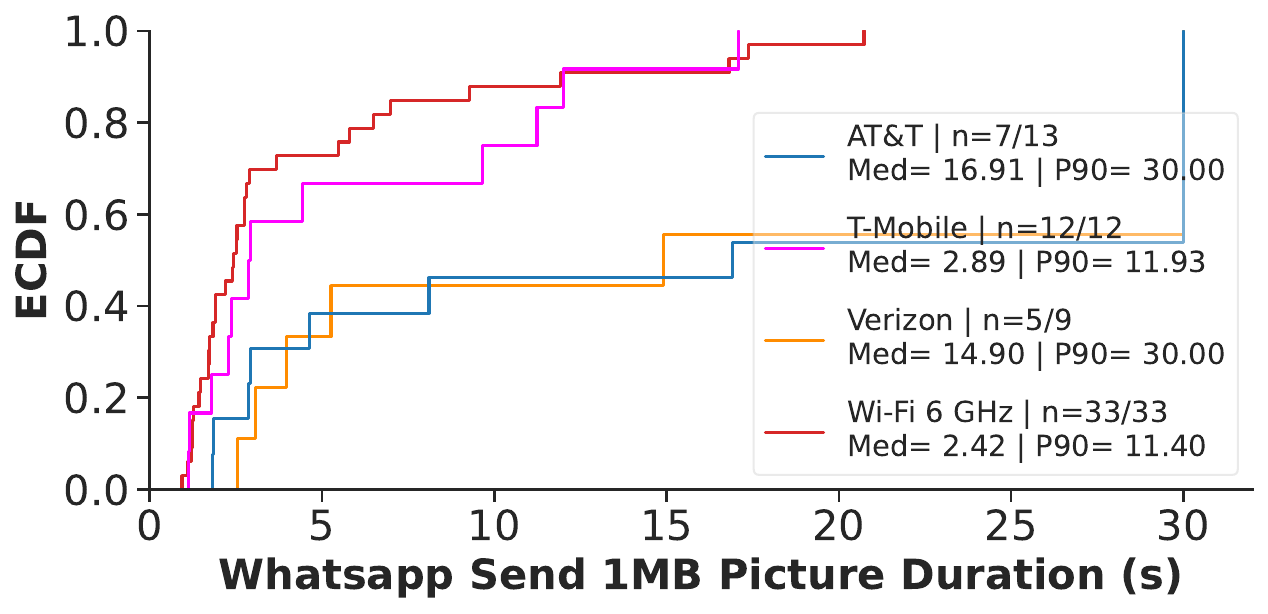}
        \vspace{-1.5em}
        \caption{WhatsApp 1~MB Image}
        \label{subfig:igwa_game_picture}
    \end{subfigure}
    \begin{subfigure}[t]{0.325\linewidth}
        \centering
        \includegraphics[width=\linewidth]{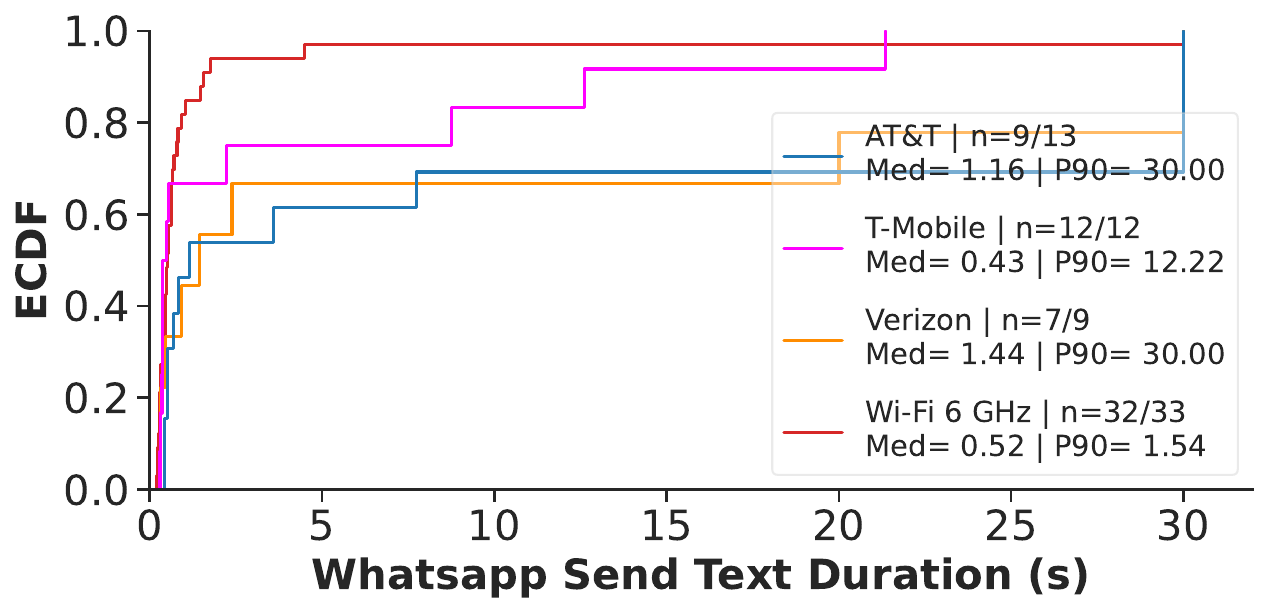}
        \vspace{-1.5em}
        \caption{WhatsApp Text}
        \label{subfig:igwa_game_text}
    \end{subfigure}
    \vspace{-.5em}
    \caption{WhatsApp Sending and Instagram Posting duration ECDF during game periods. Same data information and imputation as Fig.~\ref{fig:igwa_pregame} applies.}
    \label{fig:igwa_game}
\end{figure*}

T-Mobile benefits from its 5G SA infrastructure, which routes all browsing traffic natively and bypasses the legacy LTE small-cell DAS. This denser 5G footprint---indicated by the high unique PCI count in Table~\ref{tab:carrier_config}---proves vital, securing the lowest median TTFB at 504~ms, followed closely by the dense Wi-Fi deployment at 573~ms (Table~\ref{tab:TTFB_FST_Comparison}). Conversely, AT\&T and Verizon's reliance on EN-DC exposes latency bottlenecks inherent to shared, neutral-host LTE DAS under load, underscoring that standalone 5G is essential for stadium latency stability.
Verizon's strategy of maintaining heavy traffic over EN-DC (b66 + n77) failed under intense crowd-driven spectrum crunch, disrupting the initial handshake phase and driving its P90 TTFB to a catastrophic 5,983~ms. Meanwhile, AT\&T suffered a 36.6\% session failure rate. Its low-band n5 channel lacks the physical bandwidth to sustain such loads; when the scheduler aggressively forced higher-order downlink modulations to maximize spectral efficiency (Fig.~\ref{fig:MCS-violin}), it triggered a median BLER of 70\%. The resulting link-layer retransmissions caused extensive page-load timeouts, mirroring the carrier's underperformance in Speedtest downlink throughput benchmarks ($\S$~\ref{sec:ookla_st}). This initial packet starvation is confirmed by Table~\ref{tab:TTFB_FST_Comparison}: AT\&T's median FST collapsed from 1179~KB/s down to 298~KB/s, with a P10 capacity floor of just 121~KB/s during the critical first second of transmission.

\subsection{Comparison of WhatsApp Sending and Instagram Posting}


For this section, we compare the total duration from transmission initiation to successful receipt (for WhatsApp text and images) and from upload initiation to completion (for Instagram posts) as illustrated in Fig.~\ref{fig:igwa_pregame} and Fig.~\ref{fig:igwa_game}. Since a 30.00~s timeout policy was enforced in our QualiPoc test definition, we perform data imputation by assigning a uniform value of 30.00~s for failed tests that exceeds the timeout. This ensures the systemic impact of dropped connections or scheduling timeouts under peak load is preserved within the high-percentile distributions.

While non-game periods show near-100\% session completion across all networks, game-day stadium load triggers severe failure rates. Using Wi-Fi 6~GHz as a baseline, which experiences minimal failures in text (3.0\%) and images (0.0\%), alongside a moderate 14.7\% failure rate for Instagram posting, cellular performance diverges drastically. T-Mobile matches Wi-Fi resilience with a 0.0\% failure rate for both text and images, though it drops to a 23.1\% failure rate for Instagram posts. In contrast, AT\&T and Verizon exhibit catastrophic drops: text failures rise to 30.8\% (AT\&T) and 22.2\% (Verizon), while WhatsApp image failures climb to 46.2\% and 44.4\%, respectively. For Instagram posting, Verizon matches the highest failure rate at 44.4\%, while AT\&T sits at 30.8\%. This congestion similarly compromises user-perceived median latencies. Compared to non-game baselines, median text degradation remains relatively controlled, scaling by just 1.3$\times$ on T-Mobile, 1.4$\times$ on Wi-Fi, 4.0$\times$ on AT\&T, and 4.5$\times$ on Verizon (Fig.~\ref{subfig:igwa_pregame_text} and Fig.~\ref{subfig:igwa_game_text}). However, for 1~MB payloads, AT\&T and Verizon suffer the worst degradation of all networks. For WhatsApp images (Fig.~\ref{subfig:igwa_pregame_picture} and Fig.~\ref{subfig:igwa_game_picture}), AT\&T's median latency expands by an acute 11.7$\times$ (from 1.44~s to 16.91~s) and Verizon's by 10.6$\times$ (from 1.40~s to 14.90~s), whereas T-Mobile and Wi-Fi stay highly stable, degrading by only 2.1$\times$ and 1.8$\times$. For Instagram posts (Fig.~\ref{subfig:igwa_pregame_posting} and Fig.~\ref{subfig:igwa_game_posting}), Verizon also registers the worst median inflation at 2.1$\times$.

Evaluating absolute performance on game (Fig.~\ref{fig:igwa_game}) reveals a latency hierarchy. Wi-Fi 6~GHz delivers the best baseline experience, holding the high-percentile text tail to just 1.54~s (P90) and maintaining a 2.42~s median for images (Fig.~\ref{subfig:igwa_game_picture}). Among cellular networks, T-Mobile significantly outperforms its peers, emerging as the best cellular operator. It is the only carrier to bypass the 30.00~s tail at P90 for both text (12.22~s) in Fig.\ref{subfig:igwa_game_text} and images (11.93~s) in Fig.\ref{subfig:igwa_game_picture}, tracking close to Wi-Fi performance. This resilience can be attributed to T-Mobile's well-performing 5G SA (n41 and n25), where all connections during these tests went to. Conversely, AT\&T and Verizon struggle severely across all payload profiles under stadium load. At high percentiles, Fig.~\ref{fig:igwa_game} shows both operators collapse into the 30.00~s timeout ceiling at P90 for small text messages, large images, and Instagram posts. The physical layer analysis shows a consistent pattern with $\S$\ref{sec:ookla_st} where the uplink traffic is routed to LTE low-bands because of their better channel quality. This suggests that the routing affects latency across different packet sizes. This demonstrates that small-packet scheduling on AT\&T and Verizon also breaks down entirely under high user density.

\section{Conclusions \& future research} \label{sec_conclusions}

In this work, we presented a comprehensive, data-driven empirical analysis of user-facing Quality of Experience (QoE) and network-layer performance within an ultra-dense environment, benchmarking multi-carrier 4G/5G cellular systems against a dense 5/6~GHz Wi-Fi deployment during sold-out games at Notre Dame Stadium. By capturing both network-layer metrics and active application-layer interactions, our findings expose critical operational realities of modern wireless infrastructure under extreme crowd density.

Our measurement campaign empirically confirms a severe, persistent ``uplink gap`` under extreme crowd densities, demonstrating that stadium-scale performance is fundamentally dictated by architectural paradigms and spectrum constraints. Across all benchmarks, the dense stadium Wi-Fi deployment delivers the most robust experience that leverage deployment densification, as shown by downlink throughput performance comparable to the best performing T-Mobile networks, as well as excellent uplink throughput and latency performances: lowest page-load failure rate of 3.9\% during games and low degradation of WhatsApp and Instagram test up to 2.1$\times$ compared to non-games.
Among cellular operators, T-Mobile emerges as overall best performer with its native 5G SA architecture, leveraging its mid-band spectrum to achieve the lowest game-day median TTFB of 504~ms, closely rivaling Wi-Fi's 573~ms. Conversely, AT\&T and Verizon experience severe degradation due to their EN-DC architectures. Relying on a shared LTE anchor exposes both networks to massive scheduling and handshake congestion, forcing a catastrophic collapse in median downlink MCS (down to 6.0 for AT\&T and 4.9 for Verizon).
This architectural bottleneck is further exacerbated by clear spectral trade-offs. While 4G FDD bands showed better channel quality during uplink throughput tests, AT\&T utilized narrow low-band FDD n5 that demonstrated an extreme 70\% median BLER under browsing test, leading to a 36.6\% session failure rate and FST drop to 298~KB/s. Meanwhile, Verizon's mid-band n77 deployment struggled with acute path loss (-106~dBm median NR RSRP), causing P90 TTFB degradation to 5,983~ms. Ultimately, these results demonstrate that a mature, standalone 5G deployment or a robust Wi-Fi offload strategy is essential to prevent routine application tasks from succumbing to severe tail latencies under extreme stadium loads.

To scale future ultra-dense deployments, cellular networks require dynamic TDD uplink scheduling and low-band densification to alleviate small-packet starvation. Concurrently, highly distributed unlicensed 5/6~GHz Wi-Fi should serve as a primary offload layer to absorb localized capacity strain. Empirically evaluating emerging features designed to close this ``uplink gap``---including 5G uplink carrier aggregation and uplink transmit switching, alongside Wi-Fi~7 Multi-Link Operation and preamble puncturing---remains a critical direction for future work.



\bibliographystyle{IEEEtran}
\bibliography{main}

\end{document}